\documentclass[aps,prl,twocolumn,superscriptaddress]{revtex4}
\usepackage{subfigure}
\usepackage{graphicx}
\usepackage{rotating} \usepackage{color}
 \usepackage{amsmath,amsthm}
\usepackage{epstopdf}
\begin{document}
\title{Evidence of disorder in biological molecules from single molecule pulling experiments}
\author{Changbong Hyeon}
\thanks{hyeoncb@kias.re.kr}
\affiliation{Korea Institute for Advanced Study, Seoul 130-722, Korea}
\author{Michael Hinczewski}
\author{D. Thirumalai}
\thanks{thirum@umd.edu}
\affiliation{Institute for Physical Science and Technology, University of Maryland, College Park, Maryland 20742, USA}

\begin{abstract}
Heterogeneity in biological molecules, resulting in molecule-to-molecule variations in their dynamics and function, is an emerging theme.  
To elucidate the consequences of heterogeneous behavior at the single molecule level, we propose an exactly solvable model in which the unfolding rate due to mechanical force depends parametrically on an auxiliary variable representing an entropy barrier arising from fluctuations in internal dynamics. 
When the rate of fluctuations $-$ a measure of dynamical disorder $-$ is comparable to or smaller than the rate of force-induced unbinding, we show that there are two experimentally observable consequences: non-exponential survival probability at constant force, and a heavy-tailed rupture force distribution at constant loading rate.  
By fitting our analytical expressions to data from single molecule pulling experiments on proteins and DNA, we quantify the extent of  disorder.  
We show that only by analyzing data over a wide range of forces and loading rates can the role of disorder due to internal dynamics be quantitatively assessed.
\end{abstract}
\maketitle

% Recent single-molecule experiments on biomolecules have underscored
% molecular heterogeneities resulting from unusually slow dynamics of
% individual molecules \cite{Hyeon2012NatureChem,Liu2013Nature}.
% Manifestation of such heterogeneity (or quenched disorder) becomes
% more plausible as size of a system grows, yet its molecular origin
% in dynamics of small biomolecules is difficult to identify.
% Quantitative analysis on individual time traces of Holliday
% junctions revealed that several noninterconvertible patterns of
% single molecule data is due to a number of distinct conformational
% substates pinned by Mg$^{2+}$ ions \cite{Hyeon2012NatureChem}.
Complex systems, characterized by processes that occur over a wide spectrum of time and length scales, often exhibit heterogeneous
behavior.  Spin glasses with quenched randomness \cite{Mezardbook} and structural glasses in which randomness is self-generated
\cite{Kirkpatrick89JPhysA,Parisi2010RMP} are two classic examples where heterogeneity is indicated by the violation of the law of
large numbers \cite{ThirumalaiPRA89}.  
These systems exhibit sub-sample to sub-sample variations in measurable quantities.
For biological systems, it is increasingly becoming appreciated that there are cell-to-cell variations (on length scales $\sim$ $\mu$m) \cite{altschuler2010cell,Pelkmans12Science,Gross13CurrBiol} as well as molecule-to-molecule variations ($\sim$ nm scales).
Manifestation of heterogeneous behavior on the $\mu$m length scale is easier to fathom than on the molecular scale. In a pioneering study, evidence for disorder in enzymes was presented using single molecule experiments \cite{Xie98Science}.
Two recent studies \cite{Hyeon2012NatureChem,Liu2013Nature} have further established that on the nm scale biological molecules
display heterogeneity and broken ergodicity just as found in cells \cite{Weitz11PNAS} and glasses \cite{Biroli2013JCP}.  Time traces
generated using single molecule fluorescence energy transfer experiments on the Holliday junction, a mobile junction of four
DNA strands involved in exchange of genetic information, showed that the conformational space is partitioned into disjoint basins of
attraction \cite{Hyeon2012NatureChem}.  Interconversions between the basins do not occur on long time scales unless the system is
annealed by first reducing the concentration of Mg$^{2+}$ ions for a period of time, and then increasing Mg$^{2+}$.  More recently, it
has been demonstrated that the speed of the DNA unwinding motor RecBCD varies from one molecule to another.  
Persistent heterogeneity in speed can be ``reset" by an annealing protocol, which involves depleting and then reintroducing Mg$^{2+}$-ATP to the enzyme \cite{Liu2013Nature}. 
Both these experiments on unrelated systems show that there must be intrinsic disorder between chemically
identical molecules.

If the dynamical variations from molecule-to-molecule in the Holliday junction and RecBCD helicase are due to disorder,
then it should be possible to discern the consequences in single molecule pulling experiments, which probe the response of proteins and
nucleic acids to mechanical force. Previously, such experiments have been particularly useful in directly measuring
some features of the folding landscape \cite{Hyeon03PNAS,Reich05EMBOrep,lannon2012BJ,Kuo10PNAS} of biological molecules that are difficult to access by other methods.
Here, we show two signatures of molecular disorder: deviations from exponential kinetics in force-induced unfolding of proteins
\cite{lannon2012BJ,Kuo10PNAS}, and the presence of fat tails in the distribution of rupture forces, $P(f)$,
characterizing the unzipping of DNA \cite{Strunz99PNAS}. 
These features cannot be explained using standard theories, which involve crossing a one-dimensional free energy barrier in the presence of force.  
Instead, we propose a generic mechanism, based on a model coupling molecular disorder and function.  
As an illustration, consider the unbinding (or binding) kinetics of a ligand from a receptor molecule, where the dynamics depends on the time-varying conformation of the receptor (open or closed)~\cite{Zwanzig92JCP}.  
The ligand is more tightly bound in the closed than in the open conformation. 
Depending on the gating rate $\lambda$ (the frequency of transitions between the conformations) the ligand is expected to exhibit very
different unbinding kinetics.  If $k$ is the mean rate of unbinding, and $k/\lambda\gg 1$ or $k/\lambda\ll1$, the environment
appears \emph{static} to the ligand \cite{Zwanzig92JCP,Zwanzig90ACR}.
The ligand experiences either \emph{quenched disorder} ($k/\lambda\gg 1$), unbinding via parallel paths over a spectrum of multiple barriers, or \emph{annealed disorder} ($k/\lambda\ll1$), unbinding via a single path over a rapidly averaging barrier.  
If $k/\lambda\sim \mathcal{O}(1)$, the gating produces a fluctuating environment along the dynamic pathway of the ligand and affects the unbinding process in a non-trivial fashion. This regime is often termed \emph{dynamic disorder} \cite{Zwanzig92JCP,Zwanzig90ACR}.  
The gating mechanism has been extensively studied in both experiments and theories in the context of oxygen binding to myoglobin \cite{Yue80Biochem,Frauenfelder01PNAS,Zwanzig92JCP}.  
The presence of dynamical disorder in the oxygen-myoglobin system results in a power law decay of unreacted oxygen
%at short time ($\tau\ll\lambda^{-1}$) and exponential decay at long time scale ($\tau\gg \lambda^{-1}$), 
and a fractional order dependence of binding rate constant on solvent
viscosity \cite{Yue80Biochem}.  
To account for the origin of this phenomenon, Zwanzig proposed a fluctuating bottleneck (FB) model \cite{Zwanzig92JCP}, which considers a rate process controlled by passage through a bottleneck whose cross-sectional area, responsible
for the reactivity, undergoes stochastic fluctuations.

%\begin{figure}[ht]
%\includegraphics[width=3.40in]{Fig1.eps}
%\caption{Fluctuating bottleneck model. A. Oxygen binding pocket of myoglobin that gates the kinetics of oxygen with a frequency $\lambda$ B. Fluctuations of $r$-coordinate with varying $\lambda$ values. 
%The autocorrelation of $r$-coordinate is given as $\langle r(t+\tau)r(t)\rangle=\langle r(\tau)r(0)\rangle=\langle r^2\rangle e^{-\lambda\tau}$. \label{Fig1}}
%\end{figure}

While the frequency $\lambda$ governing the internal dynamics, which
is intrinsic to a molecule, can in principle be varied to a certain
extent by changing viscosity \cite{Yue80Biochem}, the unbinding rate
$k$ can be more easily altered by changing force $f$ or
  the corresponding most probable force $f^*$ under the constant
  loading rate condition in single-molecule pulling experiments, thus
providing a way to infer dynamic disorder.  Here, we adopt Zwanzig's
FB concept as a general mechanism for probing the internal disorder in
biological molecules, with explicit experimental consequences.  By
fitting our analytical expressions to single-molecule force data, we
extract a measure of dynamic disoder in proteins and DNA.

%{\bf Theory.} 
To model the effect of mechanical force on the dynamics of crossing a free energy barrier in the presence of molecular gating,
we modified Zwanzig's FB model \cite{Zwanzig92JCP} using an effective potential $U_\text{eff}(x;r)=U(x;r)-fx$ that depends parametrically on $r$, the auxiliary variable characterizing the internal dynamics, and explicitly on the molecular extension $x$ conjugate to the applied force, $f$ \cite{Hyeon07JP}. 
The FB model is governed by two Langevin equations of motion:
\begin{align}
\zeta\partial_t x&=-\partial_x U_\text{eff}(x;r)+F_x(t)\nonumber\\
\partial_t r&=-\lambda r+F_r(t)
\label{eqn:FB_model}
\end{align}
where $\zeta$ is the friction coefficient along $x$. 
The precise functional form of $U(x;r)$ is arbitrary except it should have a local minimum corresponding to a bound (folded) state at
  $x=x_\text{b}$, separated by a free energy barrier at $x=x_\text{ts}>x_\text{b}$ from the unbound (unfolded) ensemble at large $x$.
The variable $r$ is the dimensionless bottleneck radius, imposed with a reflecting boundary condition at $r=0$ to satisfy $r\geq 0$ \cite{Zwanzig92JCP}.
Both the noise-related random force $F_x(t)$ along $x$ and $F_r(t)$, the stochastic fluctuation of $r$, satisfy the fluctuation-dissipation theorem: 
$\langle F_x(t)F_x(t')\rangle=2\zeta k_BT\delta(t-t')$ and $\langle  F_r(t)F_r(t')\rangle=2\lambda\theta\delta(t-t')$, with $k_BT$ being
  the thermal energy, and $\langle r^2\rangle\equiv \theta$.  
  Forced-unbinding occurs on first passage from $x_\text{b}$ to $x_\text{ts}$, with a rate $K(f,r)$ that in general
  varies with both $f$ and $r$.  
  In traditional models of barrier crossing, there is no coupling between reaction dynamics in $x$ and other degrees of freedom, so $K$ only depends on $f$.  
  For example, in the Bell approximation $K(f) \propto e^{f\Delta x^{\ddagger}/k_BT}$, where $\Delta x^\ddagger =
  x_\text{ts}-x_\text{b}$.  In the FB model, the coupling to $r$ is incorporated by making the reaction sink proportional to the area of the bottleneck, $K(f,r) \equiv k(f) r^2$.  
The form of $K(f,r)$ is physical for the applications here because the rate of unfolding of proteins or unzipping of DNA should increase as the solvent accessible area ($\propto r^2$) increases. 
  For simplicity, we assume the
  force-dependence is described by the Bell approximation, $k(f) = k_0
  e^{f\Delta x^{\ddagger}/k_BT}$, though the calculations below can be
   generalized to more complicated models where $k(f)$ reflects
  movement of the transition state under force \cite{Hyeon07JP,Dudko06PRL}.  The Langevin
equations in Eq.\ref{eqn:FB_model} can be translated into the following Smoluchowski
equation (see Supporting Information (SI) for details):
\begin{equation}
\partial_t\overline{C}(r,t)=\left[\mathcal{L}_r(r)-k(f)r^2\right]\overline{C}(r,t) 
\label{eqn:step2}
\end{equation}
where $\overline{C}(r,t)$ is the mean probability of finding the
  system still bound ($x<x_\text{ts}$) with bottleneck value $r$ at
  time $t$, and $\mathcal{L}_r(r)=\lambda\theta\partial_r\left(\partial_r+r/\theta\right)$ \cite{Zwanzig92JCP,Hyeon07JP}.  Depending on whether $f$ is constant
  or is a linearly varying quantity with time, i.e.,
  $f(t)=\gamma t$, our problem is classified into unbinding
  under force-clamp or force-ramp conditions, respectively.

%{\it Survival probability under force-clamp condition :} 
  {\it Force-clamp}: For a constant $f$, Eq.~\ref{eqn:step2} for $\overline{C}(r,t)$ is solved analytically with a reflecting
  boundary condition at $r=0$, and an initial condition $\overline{C}(r,0)=\sqrt{\frac{2}{\pi\theta}} e^{-r^2/2\theta}$ with $r\geq 0$.
  The resulting survival probability  $\Sigma^{f}_{\lambda}(t)=\int^{\infty}_0dr\overline{C}(r,t)$ is an extension of the result by Zwanzig \cite{Zwanzig92JCP} in the presence of force, $f$ \cite{Hyeon07JP}:
\begin{align}
\Sigma^{f}_{\lambda}(t)=e^{-\frac{\lambda}{2}(S(f)-1)t}\left[\frac{(S(f)+1)^2-(S(f)-1)^2E}{4S(f)}\right]^{-1/2}
\label{eqn:solution}
\end{align}
where $S(f)\equiv \left(1+\frac{4k(f)\theta}{\lambda}\right)^{1/2}$ and $E\equiv e^{-2\lambda S(f)t}$.  In two asymptotic limits of
  $\lambda$, the expression for $\Sigma_{\lambda}^f(t)$ becomes simple.  (i) For $4k(f)\theta/\lambda\ll 1$, we have $S \approx
  1$ and the survival probability decays exponentially, $\Sigma^{f}_{\lambda}(t)=\exp{\left(-k(f)\theta t\right)}$, with
  $k(f)\theta$ acting as an effective rate constant.  (ii) For $4k(f)\theta/\lambda\gg1$, we get $S \gg 1$ and the survival
probability exhibits a power-law decay, $\Sigma^{f}_{\lambda}(t)=(1+2k(f)\theta t)^{-1/2}$ at short times
$t \ll (k(f)\theta\lambda)^{-1/2}$, changing over into an exponential decay with rate $k(f)\theta$ at long times $t \gg (k(f)\theta\lambda)^{-1/2}$.  
In the limit of quenched disorder, as $\lambda \to 0$, the power-law decay extends to all times.

Using the Bell force dependence for $k(f)$ in Eq.~\ref{eqn:solution}, we find
$S=\left(1+e^{\Lambda(f)}\right)^{1/2}$, where $\Lambda(f)=\frac{\Delta x^{\ddagger}}{k_BT}(f-f_\text{cr})$ with
$f_\text{cr}=\frac{k_BT}{\Delta x^{\ddagger}}\log{\left(\frac{\lambda}{4k_0\theta}\right)}$.
Two limiting conditions arise: (1) For $\lambda \gg 4k_0\theta$, $\Lambda(f)$ changes sign from negative to positive at $f=f_\text{cr}$. 
Therefore, as $f$ is increased, a crossover occurs from an exponential kinetics with $S \approx 1$ ($f<f_\text{cr}$) to a power-law
kinetics with $S \gg 1$ ($f>f_\text{cr}$).  
(2) For $\lambda \ll4k_0\theta$, $\Lambda(f)\gg 1$ and $S \gg 1$; hence the power-law behavior persists at all $f$. 
%We will show that unfolding experiments of polyubiquitin \cite{Fernandez06NaturePhysics,Kuo10PNAS} belongs to this category.  

{\it Force-ramp:} In constant loading rate experiments, where the
external force $f=\gamma t$ is ramped at a fixed rate
$\gamma$, $k(f)$ in Eq.~\ref{eqn:step2} is time-dependent, i.e.,
$k(f)=k[f(t)]=k_0e^{\tilde{\gamma} t}$ where
$\tilde{\gamma}=\gamma\Delta x^{\ddagger}/k_BT$.  Thus,
$\overline{C}(r,t)$ satisfies a Smoluchowski equation with a
time-dependent sink,
$\mathcal{S}(r,t)=k_0r^2e^{\tilde{\gamma}t}$, with an
initial condition
$\overline{C}(r,0)=\sqrt{2/\pi\theta}e^{-r^2/2\theta}$, and a
reflecting boundary condition $\partial_r\overline{C}(0,t)=0$.  The
survival probability at time $t$ in this case can be analytically
computed (see SI for details of derivation), leading to the result
\begin{align}
\Sigma^{\gamma}_{\lambda}(t)
=\sqrt{2}e^{\frac{\lambda t}{2}}\left[\frac{\mathcal{I}(\rho)}{\mathcal{I}(\rho_0)}\right]^{-1/2}\left[1+\kappa(t)\frac{\mathcal{I}'(\rho)}{\mathcal{I}(\rho)}\right]^{-1/2},
\label{eqn:surv}
\end{align}
where $\rho\equiv \beta\kappa(t)$, $\beta\equiv
  2\lambda/\tilde{\gamma}$, $\kappa(t)\equiv
  \sqrt{\frac{4k_0\theta}{\lambda}}e^{t\tilde{\gamma}/2}$,
  $\rho_0\equiv\rho(0)$, and $\mathcal{I}(\rho)\equiv
\left(I_{\beta}'(\rho_0)\mathcal{Q}_{\beta}(\rho)-\mathcal{Q}_{\beta}'(\rho_0)I_{\beta}(\rho)\right)-[\kappa(0)]^{-1}\left\{I_{\beta}(\rho_0)\mathcal{Q}_{\beta}(\rho)-\mathcal{Q}_{\beta}(\rho_0)I_{\beta}(\rho)\right\}$.
Here $I_{\beta}(\rho)$ is a modified Bessel function of the 1st
kind, $\mathcal{Q}_{\beta}(\rho)=I_{-\beta}(\rho)$ when $\beta$ is not
an integer, and $\mathcal{Q}_{\beta}(\rho)=K_{\beta}(\rho)$, when
  $\beta$ is an integer, where $K_{\beta}(\rho)$ is a modified Bessel
function of the 2nd kind.  An analytical but rather complicated
expression for the rupture force distribution
  $P_{\lambda}^{\gamma}(f)$ is obtained from
$P_\lambda^\gamma(f)=-\gamma^{-1}d\Sigma_{\lambda}^{\gamma}(t)/dt$
(see SI).  As will be discussed below, the distributions
  $P_{\lambda}^{\gamma}(f)$ for finite $\lambda$ exhibit fat tails
  at large $f$, a consequence of disorder.  By fitting the
theoretical expression for $P_{\lambda}^\gamma(f)$ (Eq.~S19 in the SI)
to measured force distribution data, one should be able to quantify
the extent of disorder in terms of $\lambda$.

In the two limits of $\lambda\rightarrow \infty$ and $0$, one can
  obtain explicit expressions for the most probable rupture force
  $f^*$ as a function of $\tilde{\gamma}$: for $\lambda \to \infty$,
  $f^*=(k_B T/\Delta
  x^\ddagger)\log{\left[\tilde{\gamma}/(k_0\theta)\right]}$, while for
  $\lambda \to 0$, $f^*=(k_B T/\Delta
  x^\ddagger)\log{\left\{\left(\tilde{\gamma}/k_0\theta\right)\left(1-2k_0\theta/\tilde{\gamma}\right)\right\}}$
  (see SI).  The latter equation is valid when
  $\tilde{\gamma}\geq 3k_0\theta$.  For $\lambda \to 0$ and
  $0<\tilde{\gamma}<3k_0\theta$, $P^\gamma_\lambda(f)$ is peaked at
$f=0$.  The presence of finite probability at $f=0$ is due to
unfolding through spontaneous transitions \cite{Evans01ARBBS}.
The difference in $f^\ast$ between the two asymptotic limits of
  $\lambda$ is maximized when $\tilde{\gamma}=3k_0\theta$ (illustrated
  in Fig.~S1E using synthetic data), but disappears for sufficiently
  high loading rates, $\tilde{\gamma}\gg 2k_0\theta$.  
  In conventional dynamic force spectroscopy (DFS) theory, $f^*$ is
linear in $\log{\gamma}$. 
However, a positive curvature often
develops when the force response of the molecule is ductile and hence the transition state location moves towards the bound
state \cite{Hyeon06BJ,Hyeon07JP,Hyeon2012JCP}. 
We predict that if a system has disorder, negative curvature should be discernible at  low loading rates in the $f^*$ versus $\log{\gamma}$ plot especially for $f^*\approx 0$ when disorder is quenched, i.e., $\lambda\rightarrow 0$ (see Fig.~S1E).  
Besides fitting $P^{\gamma}_{\lambda}(f)$ to
force distribution data, a careful investigation of $f^*$ vs
$\log{\gamma}$ plots at small $\gamma$ could be useful to capture the
fingerprints of dynamical disorder.

%To plot $P(f)$ on a graph, however, it is more practical to numerically differentiate Eq.~\ref{eqn:surv}. 
%Of note, $P(f)$ contains more detailed information of the system than $f^*$. 

%{\bf Application - Extracting the gating rate and disorder in various molecules.} 

We use our theory, derived from a single model, to analyze two
representative sets of force data, one from a force-clamp and one from a force-ramp experiment.  
  
   \begin{figure}[]
\includegraphics[width=3.00in]{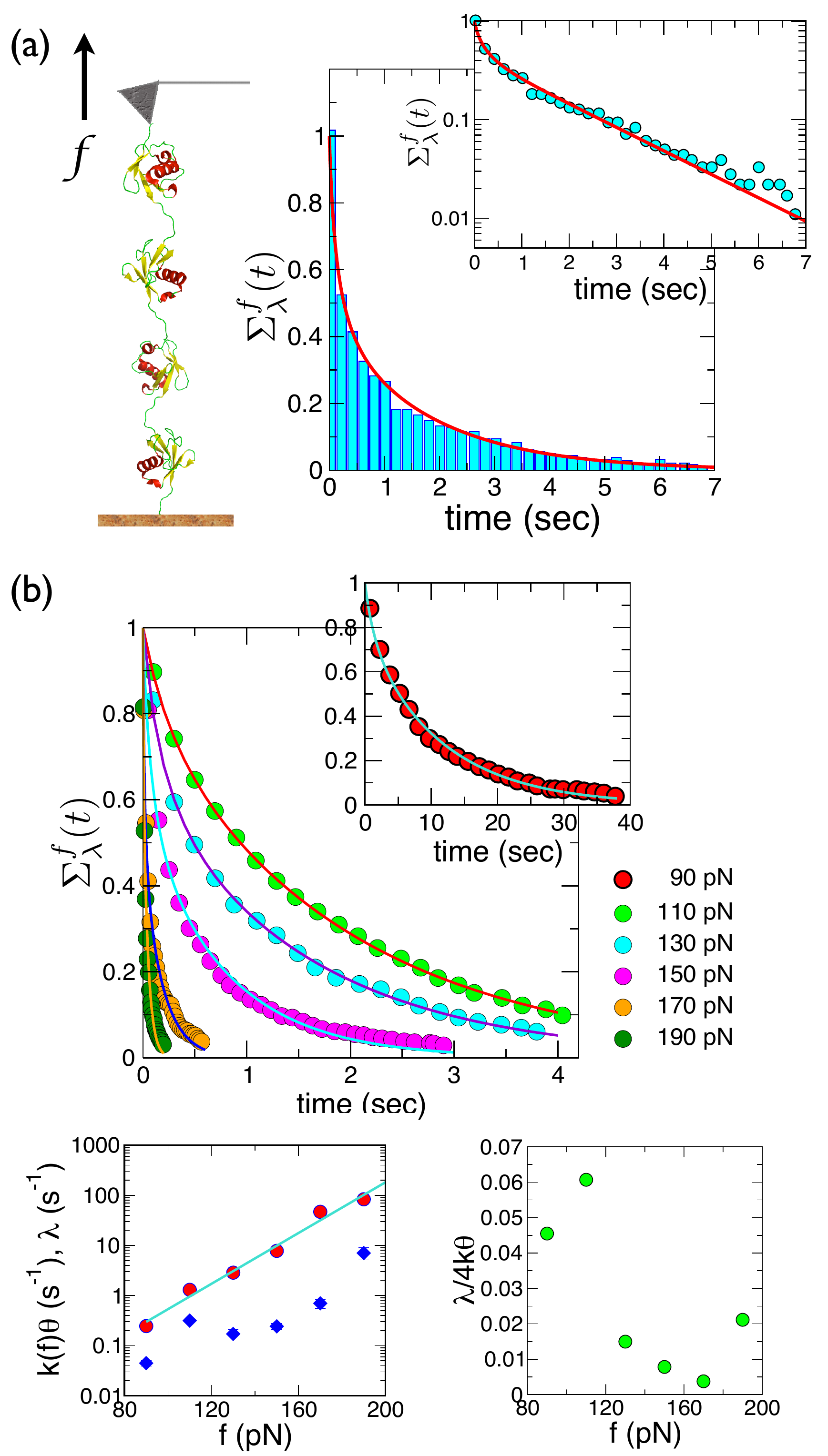}
  \caption{Interpretation of polyubiquitin data at constant force
    using the FB model.  (a) Survival probability constructed
    from dwell time analysis of polyubiquitin data in a
      force-clamp at $f=110$ pN (digitized from Fig.1 in Ref. \cite{Kuo10PNAS}). The
    line is the fit using Eq.~\ref{eqn:solution}.
    % with $k(f)\theta=5.54$ s$^{-1}$ and $\lambda=0.06$ s$^{-1}$.
    The inset shows $\Sigma^f(t)$ using a logarithmic scale.  (b)
    (top) Solid lines show our theoretical fits to the survival probability data (colored
    circles obtained by digitizing the results in Fig.2 in Ref.\cite{Kuo10PNAS}) 
    at different values of the force $f=90-190$ pN.  
    The extracted parameters $k(f)\theta$,  $\lambda$, and their ratio $\lambda/4k_0\theta$ are plotted against $f$ on the two panels at the bottom. 
  %$\log{k(f)\theta}=\log{k_0\theta}+f\Delta x^{\ddagger}/k_BT$ vs $f$ is analyzed using linear regression, which allows us to
  %determine $\Delta x^{\ddagger}=0.24$ nm and $k_0\theta=0.126$
  %\cb{s}$^{-1}$.
\label{Kuo_fit}}
\end{figure}
  
  \begin{figure}[]
\includegraphics[width=2.80in]{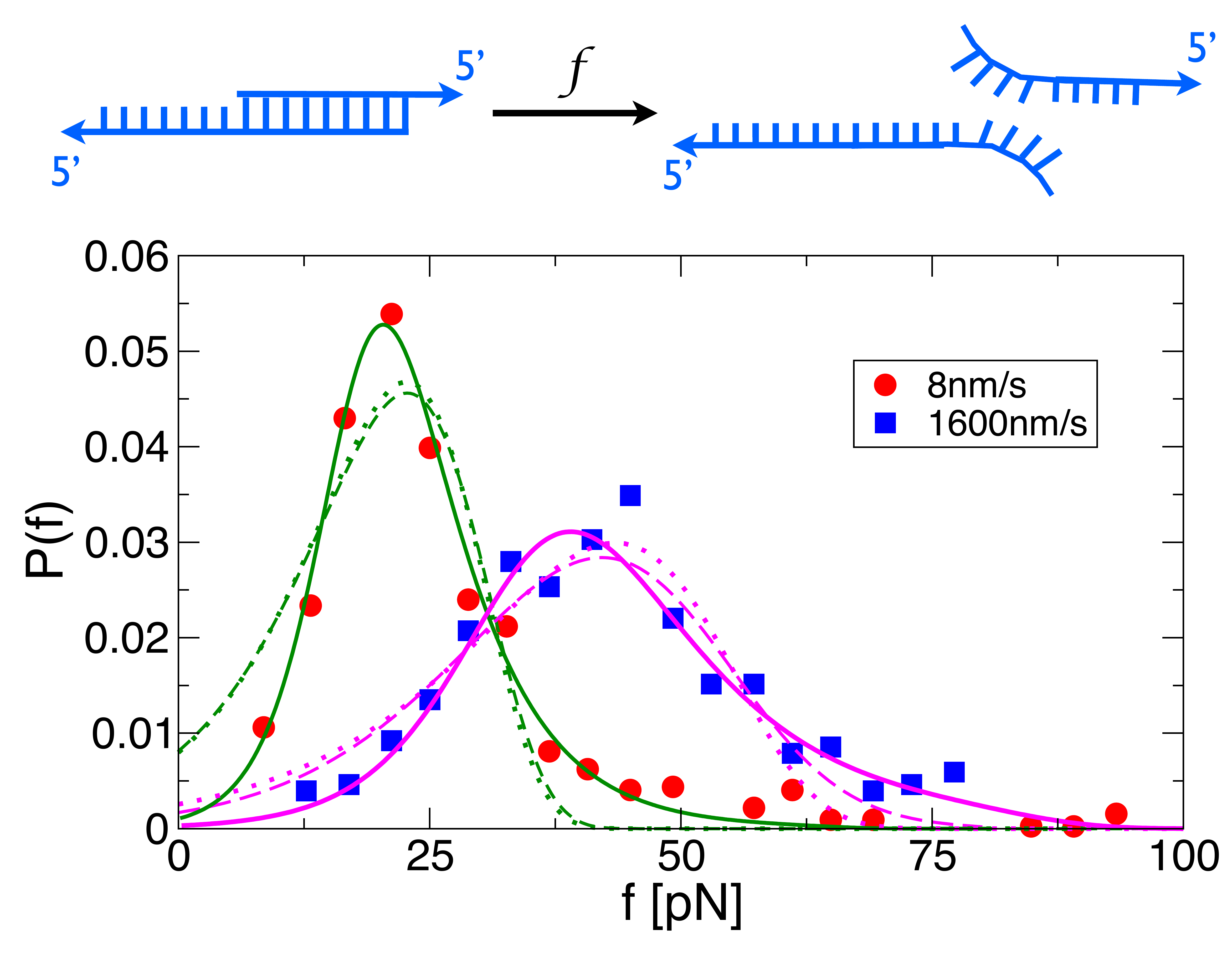}
  \caption{Analysis of rupture force distributions from
    a DNA unzipping force spectroscopy experiment (the AFM
    cantilever spring constant $\approx 2$ pN/nm)
    \cite{Strunz99PNAS} using three different models.  The fits using
    $P_{\lambda}^\gamma(f)$ (solid lines), based on our FB model, yield ($\Delta x^{\ddagger}$
    [nm], $k_0\theta$ [s$^{-1}$], $\lambda$
    [s$^{-1}$])=(1.1, 0.017, $2.8\times 10^{-5}$) for $v=8$ nm/s
    and (0.66, 0.99, 0.48) for $v=1600$ nm/s. 
    The fits using
    $P_\text{cubic}[\varepsilon(f)]=\frac{k(\varepsilon)}{\gamma}\exp{\left[\frac{k_0}{\gamma\Delta
          x^{\ddagger}}\left(1-\frac{k(\varepsilon)}{k_0}\varepsilon^{-1/2}\right)\right]}$
    (dashed lines) with $k(\varepsilon)=k_0\varepsilon^{1/2}e^{\Delta
      G^{\ddagger}(1-\varepsilon^{3/2})}$ and
    $\varepsilon(f)=1-2f\Delta x^{\ddagger}/3\Delta G^{\ddagger}$
    \cite{Dudko06PRL} yield $(\Delta x^{\ddagger}$ [nm], $k_0$
    $[s^{-1}]$, $\Delta G^{\ddagger}$ [pN$\cdot$nm])=(0.12, 0.12,
    31.9) for $v=8$ nm/s and (0.09,7.16,19.0) for $v=1600$ nm/s.  The
    fits using $P_\text{Bell}(f)=\gamma^{-1}k_0 e^{f\Delta
      x^{\ddagger}/k_BT}\exp{\left(-\gamma^{-1}k_0 k_BT/\Delta
        x^{\ddagger}(e^{f\Delta x^{\ddagger}/k_BT}-1)\right)}$ (dotted
    lines) yield $(\Delta x^{\ddagger}$ [nm], $k_0$ $[s^{-1}]$)=(0.49,
    0.13) for $v=8$ nm/s and (0.32, 8.19) for $v=1600$ nm/s.  Note
    that $P_{\lambda}(f)$ describes the heavy tails of the
    distributions better than $P_\text{cubic}(f)$ or $P_\text{Bell}(f)$,
    implying that unbinding of these DNA duplexes by force cannot be accounted
    for using one-dimensional models.
    \label{analysis}}
\end{figure}

  {\em Polyubiquitin stretching:}  
  The first is polyubiqutin stretched at constant $f$ in AFM, where the survival probabilities $\Sigma^f(t)$, obtained from dwell time analysis, exhibited non-exponential decay \cite{lannon2012BJ,Kuo10PNAS}.
The experimental data were further interpreted using a Gaussian distribution for free energy barriers and transition state locations
separating the folded and unfolded states of ubiquitin \cite{Kuo10PNAS}.  
We find that the physics of the measured non-exponential decay can be quanitatively explained using
our theory (Eq.~\ref{eqn:solution}) based on the FB model.  
The almost exact fit of the theory to experiment for $\Sigma_{\lambda}^f(t)$ at $f=110$ pN in Fig.\ref{Kuo_fit}(a) shows the presence of dynamical disorder.
% with \cb{$\lambda =0.06$ s$^{-1}$} and $k(f)\theta=5.54$ \cb{s}$^{-1}$.

For polyproteins (used in AFM experiments) there are two possible origins for disorder: (i) couplings between neighboring modules; (ii) disorder inherent to each module. Because a simple statistical relationship such as binomial factorization associated with the kinetics of individual modules \cite{Benedetti11BJ} is expected to break down for the scenario (i), experiments that change the number of modules can discriminate between the two scenarios. Regardless of the origin of heterogeneity, our theory can be used to extract the parameters characterizing disorder effects. 

If the bottleneck represents molecular gating or breathing
dynamics in a multimodular construct of polyubiquitin, 
$\lambda$ should in principle be an increasing function of $f$.  
As the applied tension increases, the amplitude of transverse fluctuations decreases, hence
  increasing the corresponding frequency.  
  Even in Bell model, the
extracted parameters $\Delta x^{\ddagger}$ and $k_0$ should be
interpreted as capturing the geometry of the landscape, which changes
with $f$ \cite{Hyeon06BJ,Hyeon07JP}.  
From the fits of
  $\Sigma(t)$ at different values of force we extracted $k(f)\theta$
  and $\lambda$ at each $f$ (bottom of Fig.\ref{Kuo_fit}(b)). The
  effective rate constant $k(f)\theta$ changes exponentially with $f$,
  in accord with Bell model.  
  By using
  $\log{k(f)\theta}=\log{k_0\theta}+\left(\frac{\Delta
      x^{\ddagger}}{k_BT}\right)f$ we obtain $\Delta
  x^{\ddagger}=0.24$ nm and $k_0\theta=0.13$ s$^{-1}$. The value
  of $\Delta x^{\ddagger}$ is in excellent agreement with experimental
  measurements \cite{Kuo10PNAS}.
  Interestingly, we also find a rough exponentially
  increasing trend in $\lambda$ with $f$. 
  Increase of $\lambda$ implies that the rate of change of accessible surface increases as $f$ increases, supporting our physical
  intuition about the influence of $f$ on the internal dynamics.
 A nonexponential $\Sigma^f(t)$ in the force-clamp condition corresponds to a heavy tailed $P(f)$ in the force-ramp condition. 
 The parameters determined for polyubiquitin satisfy $\lambda/4k_0\theta\ll 1$ for all $f$ values (Fig.\ref{Kuo_fit}(b)). Therefore, a heavy tailed $P(f)$ will manifest itself over the entire range of $r_f\left(=\frac{k_0\theta k_BT}{\Delta x^{\ddagger}}e^{f^*\Delta x^{\ddagger}/k_BT}\right)\approx 230-2.4\times 10^5$ pN/s that gives rise to the most probable forces $f^*=80-200$ pN.

{\it Unbinding of DNA duplexes:} 
The effects of disorder manifest themselves dramatically in unzipping experiments on DNA duplexes between 5'-GGCTCCCTTCTACCACTGAC\underline{ATCGCAACGG}-3' and 3'-\underline{TAGCGTTGCC}-5', where underlined sequences are complimentary to each other \cite{Strunz99PNAS}.  
The measured rupture force distributions have heavy tails at high $f$, though the physical reasons for the tails were not discussed in the original paper. 
Fig.~\ref{analysis} shows the unzipping force distribution at two pulling speeds, 8 nm/s and 1600 nm/s, and fits using
our model, compared to two other models commonly employed in analyzing DFS experiments.  It is clear that our theory for
$P^{\gamma}_{\lambda}(f)$ using the FB model most accurately fits the force data. 
Two other models, based on the Bell model \cite{Evans97BJ} and a cubic potential \cite{Dudko06PRL},  fail to capture the tail part of
the data because they incorporate no disorder, only unbinding through a one-dimensional free energy profile.  For 8 nm/s,
the gating frequencies are almost zero ($\lambda\approx 2.8\times 10^{-5}$ s$^{-1}$).  However, $\lambda$ increases by nearly four
orders of magnitude to $\lambda=0.48$ s$^{-1}$ at $v=1600$ nm/s.
Given that the bubble dynamics of a DNA duplex occur with a characteristic time scale of $\sim 50$ $\mu$s
\cite{Altan-Bonnet03PRL}, our extracted $\lambda$ values are too large for breathing motion of base pairs to be a source of
disorder.  It is more reasonable to surmise that each duplex is pulled from starkly different and very slowly interconverting
conformations, similar to that found in Holliday junctions and RecBCD.  
The origin of this disorder is likely to be in the heterogeneity of base pairings, although the experiment was intended to probe the unbinding dynamics from a homogeneous sample made of two DNA strands with perfect complimentarity.
The large increase in $\lambda$ as $v$ increases suggests that tension facilitates interconversion between states, which accords well with the expectation that force lowers barriers between distinct bound states.

Applications of our theory  reveal that by analyzing data from single-molecule pulling experiments over a range of forces and loading rates one can infer the role that dynamical disorder,
intrinsic to the molecule, plays in unfolding or unbinding kinetics.
The observed non-exponential kinetics in survival probability or
fat tails in the unfolding force distributions cannot be captured by theories based on one-dimensional free energy profiles. 
%The role
%multidimensionality plays in a system's response to force is increasingly becoming relevant, as suggested by several
%experimental studies \cite{EvansNature99,Chu09PNAS,NevoNSB03,Zhu03Nature}. 
Our work shows that by using an auxiliary coordinate in addition to extension we can quantitatively predict the consequences of disorder in the dynamics of biological molecules. The theory provides a conceptual framework for analyzing future single molecule pulling experiments on complexes involving proteins, DNA, and RNA in which heterogeneity is sure to play a prominent role.

We thank Shaon Chakraborty for useful discussions. C.H. thanks Korea Institute for Advanced Study for providing computation resources.  D.T. acknowledges a grant from the National Institutes of Health (GM 089685).

%\bibliographystyle{apsrev}
%\bibliography{mybib1}

%\clearpage

\renewcommand{\theequation}{S\arabic{equation}}
\renewcommand{\thefigure}{S\arabic{figure}}
\setcounter{equation}{0}
\setcounter{figure}{0}
\section{Supporting Information}

\noindent {\bf Derivation of the Smoluchowski equation for FB model : }
According to the Liouville theorem ($d\varphi/dt=0$) the time evolution of probability density $\varphi(x,r,t)$ in terms of $x$ and $r$ satisfies 
\begin{equation}
\frac{\partial\varphi}{\partial t}=-\frac{\partial}{\partial x}\left(\frac{dx}{dt}\varphi\right)-\frac{\partial}{\partial r}\left(\frac{dr}{dt}\varphi\right). 
\label{eqn:Liouville}
\end{equation}
Insertion of two Langevin equations (Eq.(1) in the main text) for the fluctuating bottleneck model 
$\partial_t x=-\zeta^{-1}[\partial_x U_{\text{eff}}(x;r)+F_x(t)]$ and $\partial_t r=-\lambda r+F_r(t)$ into Eq.\ref{eqn:Liouville} leads to  
\begin{widetext}
\begin{align}
\frac{\partial\varphi}{\partial t}&=\frac{\partial}{\partial x}\left(\frac{1}{\zeta}\frac{dU_{\text{eff}}(x)}{dx}\varphi\right)+\frac{\partial}{\partial r}\left(\lambda r\varphi\right)-\frac{\partial}{\partial x}\left(\frac{1}{\zeta}F_x(t)\varphi\right)-\frac{\partial}{\partial r}\left(F_r(t)\varphi\right)\nonumber\\
&\equiv -\mathcal{L}\varphi-\frac{\partial}{\partial \vec{a}}\cdot \left(\vec{F}(t)\varphi\right)
\label{eqn:Liouville2}
\end{align}
\end{widetext}
where $\vec{a}\equiv (x,r)$ and $\vec{F}(t)\equiv(\frac{1}{\zeta}F_x(t),F_r(t))$. 
Using the vector notation as in the second line of Eq.\ref{eqn:Liouville2}, one can formally solve for the probability density $\varphi(\vec{a},t)$ as 
\begin{align}
\varphi(\vec{a},t)=e^{-t\mathcal{L}}\varphi(\vec{a},0)-\int^t_0dse^{-(t-s)\mathcal{L}}\frac{\partial}{\partial \vec{a}}\cdot\left(\vec{F}(s)\varphi(\vec{a},s)\right)
\label{eqn:formal}
\end{align}
Averaging $\varphi(\vec{a},t)$ over noise after iterating $\varphi(\vec{a},t)$ into the noise related term in the integrand and exploiting the fluctuation-dissipation theorem, 
we obtain the Smoluchowski equation 
for $\varphi(x,r,t)$ in the presence of a reaction sink, $\mathcal{S}(x,r)=k_rr^2\delta(x-x_{\text{ts}})$,
\begin{equation}
\frac{\partial\overline{\varphi}(x,r,t)}{\partial t}=\left[\mathcal{L}_x(x)+\mathcal{L}_r(r)-\mathcal{S}(x,r)\right]\overline{\varphi}(x,r,t),
\label{eqn:Smol}
\end{equation} 
where $\mathcal{L}_x\equiv D\partial_x\left(\partial_x+(k_BT)^{-1}\partial_xU_{\text{eff}}(x)\right)$ and $\mathcal{L}_r\equiv \lambda\theta\partial_r\left(\partial_r+r/\theta\right)$. 
Integrating both sides of the equation over $x$ by defining $\overline{C}(r,t)\equiv\int^{\infty}_{-\infty} dx\overline{\varphi}(x,r,t)$ 
leads to $\partial_t\overline{C}=\mathcal{L}_r\overline{C}(r,t)-k_rr^2\overline{\varphi}(x_{\text{ts}},r,t)$. 
By setting 
$\overline{\varphi}(x_{\text{ts}},r,t)=\phi_x(x_{\text{ts}})\overline{C}(r,t)$ 
where $\phi(x_{\text{ts}})=e^{-U_{\text{eff}}(x_{\text{ts}})/k_BT}/\int dx e^{-U_{\text{eff}}(x)/k_BT}\approx \sqrt{U^{\prime\prime}_{\text{eff}}(x_{\text{b}})/2\pi k_BT}e^{-(U_{\text{eff}}(x_{\text{ts}})-U_{\text{eff}}(x_{\text{b}}))/k_BT}$, 
we get 
\begin{equation}
\partial_t\overline{C}(r,t)=\left[\lambda\theta\partial_r\left(\partial_r+r/\theta\right)-kr^2\right]\overline{C}(r,t), 
\label{eqn:step2}
\end{equation}
%which is Eq.(2) in the main text. 
where $k\equiv k_r\sqrt{U^{\prime\prime}_{\text{eff}}(x_{\text{b}})/2\pi k_BT}e^{-\Delta U^{\ddagger}/k_BT}$ with $\Delta U^{\ddagger}\equiv U(x_{\text{ts}})-U(x_{\text{b}})$. 
In all likelihood, $k_r\left(=D\times\sqrt{U^{\prime\prime}_{\text{eff}}(x_{\text{ts}})/2\pi k_BT}\right)$ represents the product of diffusion coefficient $D$ associated with barrier crossing dynamics and the contribution of dynamics at the barrier top. 
Thus, under tension $f$, one can set  
$k\rightarrow k(f)=k_0e^{f\Delta x^{\ddagger}/k_BT}$ where 
$k_0\equiv (\xi D\sqrt{U^{\prime\prime}_{\text{eff}}(x_{\text{b}})U^{\prime\prime}_{\text{eff}}(x_{\text{ts}})}/2\pi k_BT)e^{-\Delta U^{\ddagger}/k_BT}$ and $\xi$ describes the correction due to geometrical information of the cross section of bottleneck \cite{Zwanzig92JCP,Hyeon07JP}. 
Therefore, under tension $f$, Eq.\ref{eqn:step2} becomes Eq.(2) in the main text.\\

\noindent{\bf Solution of the Smoluchowski equation with time-dependent sink : }
For the problem with a constant loading rate, the sink function of our Smoluchowski equation becomes time-dependent, resulting in the following equation for the flux $\overline{C}(r,t)$,  
\begin{widetext}
\begin{equation}
\frac{\partial\overline{C}(r,t)}{\partial t}=\lambda\theta\frac{\partial}{\partial r}\left(\frac{\partial}{\partial r}+\frac{r}{\theta}\right)\overline{C}(r,t)-k_0r^2e^{t(\gamma\Delta x^{\ddagger}/k_BT)}\overline{C}(r,t) 
\label{eqn:PDE}
\end{equation}
with $\overline{C}(r,t=0)=\sqrt{\frac{2}{\pi\theta}}e^{-r^2/2\theta}$. 
%is analytically tractable by substituting the Zwanzig's ansatz $\overline{C}(r,t)\sim e^{\nu(t)-\mu(t)r^2}$ \cite{Zwanzig92JCP}. 
Although a time-dependent sink term, in general, makes Smoluchowski equations analytically intractable, the ansatz $\overline{C}(r,t)\sim e^{\nu(t)-\mu(t)r^2}$ used in the Ref.~\cite{Zwanzig92JCP} allows us to solve the above problem exactly.  
Substitution of $\overline{C}(r,t)\sim e^{\nu(t)-\mu(t)r^2}$ leads to two ODEs for $\nu(t)$ and $\mu(t)$ (with $'$ denoting derivative with respect to $t$), 
\begin{equation}
\nu'(t)=-2\lambda\theta\mu(t)+\lambda
\label{eqn:nu_org}
\end{equation}
and
\begin{align}
\left(\mu(t)-\frac{1}{4\theta}\right)'=-4\lambda\theta\left(\mu(t)-\frac{1}{4\theta}\right)^2+\frac{\lambda}{4\theta}\left(1+\frac{4k_0\theta}{\lambda}e^{t\tilde{\gamma}}\right)
\label{eqn:mu}
\end{align}
\end{widetext}
with $\mu(0)=1/2\theta$. 
%Note that we made the reaction sink time-dependent by changing $k$ into $k_0e^{t\tilde{\gamma}}$ from the original fluctuating bottleneck model. Nevertheless, the above differential equations are still exactly solvable. 
The equation for $\mu(t)$ in Eq.\ref{eqn:mu} is the Riccati equation, $y'=q_0(t)+q_1(t)y+q_2(t)y^2$ with $y(t)\equiv \mu(t)-1/4\theta$. 
In general, the Riccati equation can be reduced to a second order ODE. 
The variable is changed in two steps : (i) $v(t)=q_2(t)y(t)$ leads to $v'=v^2+P(t)v+Q(t)$ where $Q=q_0q_2=-\lambda^2\left(1+\frac{4k_0\theta}{\lambda}e^{t\tilde{\gamma}}\right)$ and $P=q_1+q_2'/q_2=0$. (ii) Another substitution $v(t)=-u'(t)/u(t)$ leads to $u''(t)-P(t)u'(t)+Q(t)u(t)=0$, i.e., 
\begin{equation}
u''(t)-\lambda^2\left(1+\frac{4k_0\theta}{\lambda}e^{t\tilde{\gamma}}\right)u(t)=0. 
\label{eqn:Riccati}
\end{equation}
Introducing the variable 
$\rho=\frac{2\lambda}{\tilde{\gamma}}\sqrt{\frac{4k_0\theta}{\lambda}}e^{t\tilde{\gamma}/2}=\beta \kappa(t)$ with $\beta\equiv\frac{2\lambda}{\tilde{\gamma}}$ and $\kappa(t)\equiv \sqrt{\frac{4k_0\theta}{\lambda}}e^{t\tilde{\gamma}/2}$ 
one can modify the second-order ODE in Eq.\ref{eqn:Riccati} into a more familiar modified Bessel equation,
\begin{equation}
\rho^2U_{\rho\rho}+\rho U_{\rho}-\left[\beta^2+\rho^2\right]U=0 
\label{eqn:mod_Bessel}
\end{equation}
where $u(t)=U(\rho)$. 
The solution of the above ODE is the linear combination of $I_{\pm\beta}(\rho)$ for non-integer $\beta$, and the linear combination of $I_{\beta}(\rho)$ and $K_{\beta}(\rho)$ when $\beta$ is integer. Thus, the solution of Eq.\ref{eqn:mod_Bessel} is 
\begin{equation}
U(\rho)=\left\{ \begin{array}{ll}
    c_1I_{\beta}(\rho)+c_2I_{-\beta}(\rho)& \mbox{$\beta\neq n$, $\beta>0$}\\
     c_1I_{\beta}(\rho)+c_2K_{\beta}(\rho)& \mbox{$\beta=n$ where $n=0, 1,  2\cdots $}\end{array}\right.
\end{equation}
For simplicity, we use the notation $\mathcal{Q}_{\beta}(\rho)$ to represent either $I_{-\beta}(\rho)$ or $K_{\beta}(\rho)$, 
\begin{equation}
\mathcal{Q}_{\beta}(\rho)=\left\{ \begin{array}{ll}
    I_{-\beta}(\rho)& \mbox{$\beta\neq n$, $\beta>0$}\\
    K_{\beta}(\rho)& \mbox{$\beta=n$ where $n=0,  1,  2\cdots $}\end{array}\right.
\end{equation}
Thus one obtains $\mu(t)$ using $y(t)=-\frac{u'(t)}{q_2(t)u(t)}$. 
\begin{equation}
\mu(t)=\frac{1}{4\theta}+\frac{\kappa(t)}{4\theta}
\left(\frac{I_{\beta}'(\rho)+c\mathcal{Q}_{\beta}'(\rho)}{ I_{\beta}(\rho)+c \mathcal{Q}_{\beta}(\rho)}\right).
\label{eqn:mu}
\end{equation}
Note that $I_{\beta}'(\rho)\equiv dI_{\beta}(\rho)/d\rho$.
%The derivative of $I_{\beta}(\rho)$ can be calculated using the identity
%$I_{\beta}'(\rho)=\frac{1}{2}\left[I_{\beta-1}(\rho)+I_{\beta+1}(\rho)\right]$. 
The initial condition $\mu(0)=1/2\theta$ determines the constant $c$ in Eq.\ref{eqn:mu}
\begin{equation}
c=\frac{I_{\beta}'(\rho_0)-[\kappa(0)]^{-1}I_{\beta}(\rho_0)}{[\kappa(0)]^{-1}\mathcal{Q}_{\beta}(\rho_0)-\mathcal{Q}_{\beta}'(\rho_0)}.
\end{equation}
where $\rho_0\equiv\beta \kappa(0)$. 
Thus, one obtains 
\begin{align}
\frac{\mu(t)}{\mu(0)}
%=\frac{1}{2}\left[1+\kappa(t)\frac{\left\{I'_{\beta}(\rho_0)I'_{-\beta}(p)-I'_{-\beta}(\rho_0)I'_{\beta}(p)\right\}-\gamma\left\{I_{\beta}(\rho_0)I_{-\beta}'(p)-I_{-\beta}(\rho_0)I_{\beta}'(p)\right\}}{\left\{I'_{\beta}(\rho_0)I_{-\beta}(p)-I_{-\beta}'(\rho_0)I_{\beta}(p)\right\}-\gamma\left\{I_{\beta}(\rho_0)I_{-\beta}(p)-I_{-\beta}(\rho_0)I_{\beta}(p)\right\} }\right]\nonumber\\
=\frac{1}{2}\left[1+\kappa(t)\frac{\mathcal{I}'(\rho)}{\mathcal{I}(\rho)}\right] 
\label{eqn:mu_main}
\end{align}
where 
%\begin{equation} 
%\mathcal{N}'(t)=\left\{I'_{\beta}(\rho_0)I'_{-\beta}(\rho)-I'_{-\beta}(\rho_0)I'_{\beta}(\rho)\right\}-[\kappa(0)]^{-1}\left\{I_{\beta}(\rho_0)I_{-\beta}'(\rho)-I_{-\beta}(\rho_0)I_{\beta}'(\rho)\right\}
%\end{equation}
%\begin{equation} 
%\begin{align*}
$\mathcal{I}(\rho)\equiv \left(I_{\beta}'(\rho_0)\mathcal{Q}_{\beta}(\rho)-\mathcal{Q}_{\beta}'(\rho_0)I_{\beta}(\rho)\right)-[\kappa(0)]^{-1}\left\{I_{\beta}(\rho_0)\mathcal{Q}_{\beta}(\rho)-\mathcal{Q}_{\beta}(\rho_0)I_{\beta}(\rho)\right\}$.
Recall that $\rho\equiv \beta\kappa(t)$ with $\beta\equiv 2\lambda/\tilde{\gamma}$, $\kappa(t)\equiv \sqrt{\frac{4k_0\theta}{\lambda}}e^{t\tilde{\gamma}/2}$, and $\rho_0\equiv\rho(0)$. 
Note that $\kappa(0)(\mathcal{I}'(\rho_0)/\mathcal{I}(\rho_0))=1$ is satisfied. 
%\end{equation}
Integration of Eq.\ref{eqn:nu_org} with $t$ using Eq.\ref{eqn:mu_main} and change of variable $d\rho=\frac{\beta\tilde{\gamma}}{2}\kappa(t)dt=\lambda\kappa(t)dt$
results in the expression for $\nu(t)$: 
\begin{align}
\nu(t)
=\frac{\lambda t}{2}-\frac{1}{2}\log{\left(\frac{\mathcal{I}(\rho)}{\mathcal{I}(\rho_0)}\right)}.
\label{eqn:nu}
\end{align}
With $\mu(t)$ (Eq.\ref{eqn:mu_main}) and $\nu(t)$ (Eq.\ref{eqn:nu}) in hand, we can solve 
\begin{widetext}
\begin{align}
\overline{C}(r,t)&=\sqrt{\frac{2}{\pi\theta}}\left[\frac{\mathcal{I}(\rho)}{\mathcal{I}(\rho_0)}\right]^{-1/2}\exp{\left[\frac{\lambda t}{2}-\frac{r^2}{4\theta}\left\{1+\kappa(t)\frac{\mathcal{I}'(\rho)}{\mathcal{I}(\rho)}\right\}\right]}, 
\end{align}
from which the survival probability is obtained as 
\begin{align}
\Sigma^{\gamma}_{\lambda}(t)=\int^{\infty}_0dr\overline{C}(r,t)=\frac{1}{\sqrt{2\theta}}\frac{e^{\nu(t)}}{\sqrt{\mu(t)}}=\sqrt{2}e^{\frac{\lambda t}{2}}\left[\frac{\mathcal{I}(\rho)}{\mathcal{I}(\rho_0)}\right]^{-1/2}\left[1+\kappa(t)\frac{\mathcal{I}'(\rho)}{\mathcal{I}(\rho)}\right]^{-1/2}. 
%&=\sqrt{2}e^{\frac{\lambda t}{2}}\left[\frac{\mathcal{I}(\rho)+\kappa(t)\mathcal{I}'(\rho)}{\mathcal{I}(\rho_0)}\right]^{-1/2}
\label{eqn:surv}
\end{align}
%With $\Sigma(t)$, the relation between the unbinding time distribution and survival probability, 

The $\lambda$-dependent unbinding time distribution $P_{\lambda}(t)$ are obtained from the relation 
$P_{\lambda}(t)=-d\Sigma^{\gamma}_{\lambda}(t)/dt$,
%$=k_{\lambda}(t)\Sigma^{\gamma}_{\lambda}(t)$,  
%\begin{equation}
%k_{\lambda}(t)=\frac{\lambda}{2}\left[\frac{\kappa^2(t)\mathcal{I}''(\rho)+\frac{1}{\beta}\kappa(t)\mathcal{I}'(\rho)-\mathcal{I}(\rho)}{\mathcal{I}(\rho)+\kappa(t)\mathcal{I}'(\rho)}\right]
%\label{eqn:k}
%\end{equation}
%and
\begin{equation}
P_{\lambda}(t)=\frac{\lambda e^{\lambda  t/2}}{\sqrt{2}}\left[\kappa^2(t)\frac{\mathcal{I}''(\rho)}{\mathcal{I}(\rho)}+\frac{1}{\beta}\kappa(t)\frac{\mathcal{I}'(\rho)}{\mathcal{I}(\rho)}-1\right]
\left[\frac{\mathcal{I}(\rho)}{\mathcal{I}(\rho_0)}\right]^{-1/2}\left[1+\kappa(t)\frac{\mathcal{I}'(\rho)}{\mathcal{I}(\rho)}\right]^{-3/2}.
\label{eqn:P} 
\end{equation}
\end{widetext}
Transformation to the unbinding force distribution $P_{\lambda}(\tilde{f})\left[=\tilde{\gamma}^{-1}P_{\lambda}(t)\right]$ is made through the relationship between dimensionless scaled-force ($\tilde{f}$) and time $t$: $\tilde{f}=\tilde{\gamma}t$ with $\tilde{\gamma}=\gamma\Delta x^{\ddagger}/k_BT$. \\

\begin{figure*}[]
\includegraphics[width=7.00in]{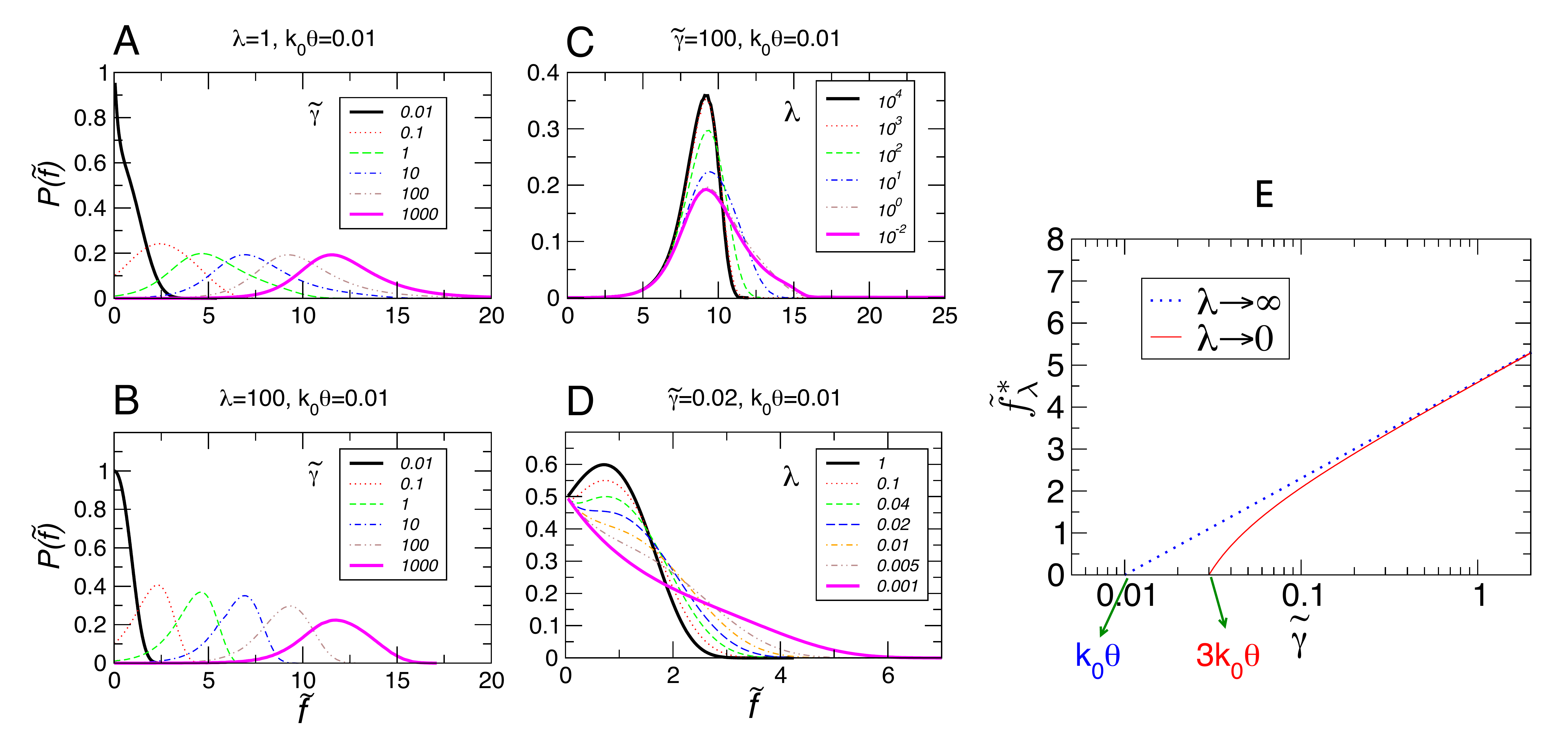}
\caption{{\bf A-D} Rupture force distributions, $P(\tilde{f})$, under varying loading rates ($\tilde{\gamma}$) and the gating frequency ($\lambda$) characterizing the disorder. {\bf E.} $\tilde{f}^*$ vs $\tilde{\gamma}$ plot under two limiting values of $\lambda$.
\label{DFS_fig}}
\end{figure*}

%The unfolding force distributions $P_{\lambda}(\tilde{f})$ for varying parameters in Fig.\ref{DFS_fig} are obtained using either Eq.\ref{eqn:P} or in practice using the numerical differentiation of $\Sigma^{\gamma}_{\lambda}(t)$. 
\noindent{\bf Illustration using synthetic data : }
Although $P_{\lambda}(\tilde{f})$ in Eq.\ref{eqn:P} is complicated, 
the familiar expression used in the Dynamic Force Spectroscopy (DFS) for $P(f)$ is restored when $\lambda\rightarrow \infty$ (see below).
In order to obtain insight into the behavior of $P_{\lambda}(\tilde{f})$ we generated several synthetic distributions for varying $\lambda$ values and loading rates. 
We find that $P_{\lambda}(\tilde{f})$ with varying $\tilde{\gamma}(=\gamma\Delta x^{\ddagger}/k_BT)$ shows the standard pattern of force distribution in DFS (Fig.\ref{DFS_fig}-A, B) \cite{Evans01ARBBS,Hyeon07JP}. 
The effect of varying $\lambda$ on $P_{\lambda}(\tilde{f})$ is shown in Fig.\ref{DFS_fig}-C, D. 
It is of particular interest that if $\tilde{\gamma}\gg k_0\theta$ then the most probable forces $f^*_{\lambda}$ from $P_{\lambda}(\tilde{f})$ are insensitive to the variation in $\lambda$ even though the shapes of $P_{\lambda\rightarrow 0}(\tilde{f})$ and $P_{\lambda\rightarrow \infty}(\tilde{f})$ are very different from each other (Fig.\ref{DFS_fig}-C). 
However, when $\tilde{\gamma}\sim k_0\theta$, $\tilde{f}^*_{\lambda}$ changes with $\lambda$ (Fig.\ref{DFS_fig}-E) and the shape of $P_{\lambda\rightarrow 0}(\tilde{f})$ differs from $P_{\lambda\rightarrow\infty}(\tilde{f})$ qualitatively (Fig.\ref{DFS_fig}-D).\\

\noindent{\bf Asymptotic behavior at $\lambda/\tilde{\gamma}\rightarrow\infty$ : } 
To obtain the asymptotic behavior we will use the following uniform asymptotic expansion of the modified Bessel function for large orders ($\nu\rightarrow\infty$) \cite{AbramowitzStegun72}. 
\begin{widetext}
\begin{align}
I_{\nu}(\nu z)&\sim \frac{1}{\sqrt{2\pi\nu}}\frac{e^{\nu\eta}}{(1+z^2)^{1/4}}\left(1+\mathcal{O}(\nu^{-1})\right)\nonumber\\
K_{\nu}(\nu z)&\sim \sqrt{\frac{\pi}{2\nu}}\frac{e^{-\nu\eta}}{(1+z^2)^{1/4}}\left(1+\mathcal{O}(\nu^{-1})\right)\nonumber\\
I_{\nu}'(\nu z)&\sim \frac{1}{\sqrt{2\pi\nu}}\frac{(1+z^2)^{1/4}}{z}e^{\nu\eta}\left(1+\mathcal{O}(\nu^{-1})\right)\nonumber\\
K_{\nu}'(\nu z)&\sim -\sqrt{\frac{\pi}{2\nu}}\frac{(1+z^2)^{1/4}}{z}e^{-\nu\eta}\left(1+\mathcal{O}(\nu^{-1})\right)
\end{align}
where $I_{\nu}'(\nu z)\equiv \frac{d}{d(\nu z)}I_{\nu}(\nu z)$ and $\eta=\sqrt{1+z^2}+\log{\left(\frac{z}{1+\sqrt{1+z^2}}\right)}$

The asymptotic behavior at large negative orders can be obtained by using the relation 
$I_{-\nu}(z)=\frac{2}{\pi}\sin{(\nu \pi)}K_{\nu}(z)+I_{\nu}(z)$
\begin{align}
I_{-\nu}(\nu z)&\sim \left(\frac{2}{\sqrt{2\pi\nu}}\sin{(\nu\pi)}\frac{e^{-\nu\eta}}{(1+z^2)^{1/4}}+\frac{1}{\sqrt{2\pi\nu}}\frac{e^{\nu\eta}}{(1+z^2)^{1/4}}\right)\left(1+\mathcal{O}(\nu^{-1})\right)\nonumber\\
I_{-\nu}'(\nu z)&\sim \left(-\frac{2}{\sqrt{2\pi\nu}}\sin{(\nu\pi)}\frac{(1+z^2)^{1/4}}{z}e^{-\nu\eta}+\frac{1}{\sqrt{2\pi\nu}}\frac{(1+z^2)^{1/4}}{z}e^{\nu\eta}\right)\left(1+\mathcal{O}(\nu^{-1})\right)
\end{align}
Using these asymptotics, we obtain the following relations at $\beta=2\lambda/\tilde{\gamma}\rightarrow \infty$. 
\begin{align}
\lim_{\beta\rightarrow\infty}\mathcal{I}(\rho)&\sim 2\left(\frac{\sin{\beta\pi}}{\beta\pi}\right)\frac{1}{\kappa(0)}\left[e^{\beta(\eta-\eta_0)}\left(1+\frac{1}{S(t)}\right)+e^{-\beta(\eta-\eta_0)}\left(1-\frac{1}{S(t)}\right)\right]\nonumber\\
\lim_{\beta\rightarrow\infty}\mathcal{I}'(\rho)&\sim 2\left(\frac{\sin{\beta\pi}}{\beta\pi}\right)\frac{1}{\kappa(0)\kappa(t)}\left[e^{\beta(\eta-\eta_0)}\left(S(t)+1\right)-e^{-\beta(\eta-\eta_0)}\left(S(t)-1\right)\right]
%\lim_{\beta\rightarrow\infty}\mathcal{I}''(\rho)&\sim 2\left(\frac{\sin{\beta\pi}}{\beta\pi}\right)\frac{1}{\kappa(0)\kappa^2(t)}\left[e^{\beta(\eta-\eta_0)}S(t)\left(S(t)+1\right)+e^{-\beta(\eta-\eta_0)}S(t)\left(S(t)-1\right)\right]
\end{align}
where $S(t)\equiv (1+\kappa^2(t))^{1/2}$.  
Therefore
%\begin{equation}
%\lim_{\beta\rightarrow \infty}\frac{\mathcal{I}''(\rho)}{\mathcal{I}(\rho)}=\frac{1+\kappa^2(t)}{\kappa^2(t)}, 
%\label{eqn:I''/I}
%\end{equation}
\begin{equation}
\lim_{\beta\rightarrow \infty}\frac{\mathcal{I}'(\rho)}{\mathcal{I}(\rho)}= \frac{S(t)}{\kappa(t)}
\left[\frac{(S(t)+1)-(S(t)-1)e^{-2\beta(\eta-\eta_0)}}{(S(t)+1)+(S(t)-1)e^{-2\beta(\eta-\eta_0)}}\right]
\label{eqn:I'/I}
\end{equation}
and 
\begin{equation}
\lim_{\beta\rightarrow \infty}\frac{\mathcal{I}(\rho)}{\mathcal{I}(\rho_0)}=e^{\beta(\eta-\eta_0)}\left[\frac{(S(t)+1)-(S(t)-1)e^{-2\beta(\eta-\eta_0)}}{2S(t)}\right]. 
\end{equation}
With $\lim_{\tilde{\gamma}\rightarrow 0}S(t)=S$ and $\lim_{\tilde{\gamma}\rightarrow0}\beta(\eta-\eta_0)
%&=\lim_{\tilde{\gamma}\rightarrow 0}\frac{2\lambda}{\tilde{\gamma}}\left[\left(1+\frac{4k\theta}{\lambda}e^{\tilde{\gamma}t}\right)^{1/2}-\left(1+\frac{4k\theta}{\lambda}\right)^{1/2}\right]+\frac{2\lambda}{\tilde{\gamma}}
%\left[\log{\frac{1+\sqrt{1+\kappa(0)^2}}{1+\sqrt{1+\kappa^2(t)}}}\right]\nonumber\\
=\lambda St$ where $S\equiv \left(1+\frac{4k_0\theta}{\lambda}\right)^{1/2}$, 
it is now easy to show
\begin{equation}
\lim_{\tilde{\gamma}\rightarrow 0}\frac{\mu(t)}{\mu(0)}=\frac{1}{2}\left[1+S\frac{(S+1)-(S-1)e^{-2\lambda St}}{(S+1)+(S-1)e^{-2\lambda St}}\right]
\label{eqn:mu_asymp}
\end{equation}
and 
\begin{equation}
\lim_{\tilde{\gamma}\rightarrow 0}\nu(t)=-\frac{\lambda t}{2}(S-1)+\log{\left[\frac{(S+1)-(S-1)e^{-2\lambda St}}{2S}\right]^{-1/2}}. 
\label{eqn:nu_asymp}
\end{equation}
Thus, substituting Eq.\ref{eqn:mu_asymp} and \ref{eqn:nu_asymp} into $\Sigma(t)=\int^{\infty}_0dr \overline{C}(r,t)=\frac{1}{\sqrt{2\theta}}\frac{e^{\nu(t)}}{\sqrt{\mu(t)}}$ recovers the previous result for survival probability in Zwanzig's FB model \cite{Zwanzig92JCP}
\begin{equation}
\lim_{\tilde{\gamma}\rightarrow 0}\Sigma (t)=\exp{\left(-\frac{\lambda}{2}(S-1)t\right)}
\left[\frac{(S+1)^2-(S-1)^2E}{4S}\right]^{-1/2}. 
\label{eqn:Survival_Zwanzig}
\end{equation}
\end{widetext}
For $\lambda\rightarrow \infty$ and $\lambda\rightarrow 0$,
$\lim_{\lambda\rightarrow\infty}\lim_{\tilde{\gamma}\rightarrow 0}\Sigma(t)=\exp{\left(-k\theta t\right)}$ and 
$\lim_{\lambda\rightarrow 0}\lim_{\tilde{\gamma}\rightarrow 0}\Sigma(t)=(1+2k\theta t)^{-1/2}$, respectively. 
\\

%\noindent{\bf Asymptotic value of $k(t)$ for $\lambda\rightarrow \infty$} 
%\begin{align}
%\lim_{\lambda\rightarrow \infty}k(t)&=\lim_{\lambda\rightarrow\infty}\frac{\lambda}{2}\left[\frac{\kappa^2(t)\mathcal{I}''(\rho)+\frac{1}{\beta}\kappa(t)\mathcal{I}'(\rho)-\mathcal{I}(\rho)}{\mathcal{I}(\rho)+\kappa(t)\mathcal{I}'(\rho)}\right]\nonumber\\
%&=\lim_{\lambda\rightarrow \infty}\frac{\lambda}{2}\left[\frac{\kappa^2(t)+\frac{1}{\beta}\kappa(t)\mathcal{I}'(\rho)/\mathcal{I}(\rho)}{1+\kappa(t)\mathcal{I}'(\rho)/\mathcal{I}(\rho)}\right]\nonumber\\
%&=\lim_{\lambda\rightarrow\infty} \frac{\lambda}{2}\left[\frac{\kappa^2(t)}{1+S(t)}\right]=k\theta e^{\tilde{\gamma}t}
%\end{align}
%where we used Eqs.\ref{eqn:I''/I} and \ref{eqn:I'/I}.\\

\noindent{\bf Survival probability ($\Sigma(\tilde{f})$) and rupture force distribution ($P(\tilde{f})$) for $\lambda\rightarrow\infty$ and $\lambda\rightarrow 0$ : }
For $\lambda\rightarrow\infty$, taking $\int^{\infty}_0dr(\cdots)$ on Eq.\ref{eqn:PDE} with pre-averaged rate constant $k(t)\theta$ and transforming $t$ into $\tilde{f}$, we obtain 
$\tilde{\gamma}\partial_{\tilde{f}}\Sigma_{\lambda\rightarrow\infty}(\tilde{f})=-k(\tilde{f})\theta\Sigma_{\lambda\rightarrow\infty}(\tilde{f})$, which leads to  
\begin{equation}
\Sigma_{\lambda\rightarrow\infty}(\tilde{f})=\exp{\left[-\frac{1}{\tilde{\gamma}}\int_0^{\tilde{f}}d\tilde{f}k(\tilde{f})\theta\right]}
\end{equation}
%where $k(\tilde{f})=k_0e^{\tilde{f}}$.
and %Using the relation %between rupture force distribution and the survival probability under tension, 
the rupture force distribution ($P(\tilde{f})=-d\Sigma(\tilde{f})/d\tilde{f}$)
%, one obtains the rupture force distribution, 
\begin{equation}
P_{\lambda\rightarrow \infty}(\tilde{f})=\frac{1}{\tilde{\gamma}}k(\tilde{f})\theta \Sigma_{\lambda\rightarrow\infty}(\tilde{f})
\label{eqn:Pf_lambda_infinity}
\end{equation}
The most probable force ($\tilde{f}^*$) is obtained using $[\partial_{\tilde{f}}P_{\lambda\rightarrow\infty}]_{\tilde{f}=\tilde{f}^*}=0$, which is equivalent to $\tilde{\gamma}[\partial_{\tilde{f}} k(\tilde{f})]_{\tilde{f}=\tilde{f}^*}=[k(\tilde{f})]^2_{\tilde{f}=\tilde{f}^*}\theta$. 
Using $k(\tilde{f})=k_0e^{\tilde{f}}$, one can easily show that 
\begin{equation}
\tilde{f}^*_{\lambda\rightarrow \infty}=\log{\left[\tilde{\gamma}/(k_0\theta)\right]}.
\label{eqn:f_lambda_infinity}
\end{equation}
This expression is equivalent to the standard DFS theory except for the presence of the $\theta$ term. 
The fast variation of $r$-coordinate effectively modifies the reactivity $k_0r^2$ into $k_0\theta$.

For $\lambda\rightarrow 0$ 
the bottleneck radius is quenched to a single value, say, $r_0$. 
In this case 
the noise-averaged probability of the molecule found at the configuration of $r_0$ at force $\tilde{f}$, $\overline{C}(r_0,\tilde{f})=\exp{\left(-\frac{1}{\tilde{\gamma}}\int^{\tilde{f}}_0d\tilde{f}k(\tilde{f})r_0^2\right)}$, should be weighted by $\phi(r_0)\left[=\sqrt{\frac{2}{\pi\theta}}e^{-r^2_0/2\theta}\right]$ as $\Sigma_{\lambda\rightarrow 0}(\tilde{f})= \int_0^{\infty}dr_0\overline{C}(r_0,\tilde{f})\phi(r_0)$ to give the survival probability,   
\begin{equation}
\Sigma_{\lambda\rightarrow 0}(\tilde{f})= \left[1+\frac{2\theta}{\tilde{\gamma}}\int^{\tilde{f}}_0d\tilde{f}k(\tilde{f})\right]^{-1/2}. 
\end{equation}
%Note that if $\tilde{\gamma}\rightarrow 0$, Zwanzig's result $\Sigma_{\lambda\rightarrow 0}(t)=(2k_0\theta t+1)^{-1/2}$ \cite{Zwanzig92JCP} is recovered. 
A similar procedure as in Eqs. \ref{eqn:Pf_lambda_infinity} and \ref{eqn:f_lambda_infinity} leads to 
%Using $P(\tilde{f})=-d\Sigma(\tilde{f})/d\tilde{f}$, one obtains rupture force distribution for $\lambda\rightarrow 0$
\begin{equation}
P_{\lambda\rightarrow 0}(\tilde{f})=\frac{1}{\tilde{\gamma}}k(\tilde{f})\theta\left[\Sigma_{\lambda\rightarrow 0}(\tilde{f})\right]^3
\end{equation}
and
%$dP_{\lambda\rightarrow 0}(\tilde{f})/d\tilde{f}|_{\tilde{f}=\tilde{f}^*}=0$ leads to  
\begin{equation}
\tilde{f}^*_{\lambda\rightarrow 0}=\log{\left\{\left(\tilde{\gamma}/k_0\theta\right)\left(1-2k_0\theta/\tilde{\gamma}\right)\right\}}. 
\label{eqn:smalllambda}
\end{equation}
\\

\noindent{\bf Comparison between $P_{\lambda\rightarrow\infty}(\tilde{f})$ and $P_{\lambda\rightarrow 0}(\tilde{f})$ one-dimensional models :} 
Asymptotic behaviors of $P(f)$ with two limiting $\lambda$ values at large $\tilde{f}\gg\tilde{f}^*$, $P_{\lambda\rightarrow\infty}(\tilde{f})$ and $P_{\lambda\rightarrow 0}(\tilde{f})$ are obtained by using the Bell model for $k(\tilde{f})$. 
Comparison between $P_{\lambda\rightarrow\infty}(\tilde{f})$ and $P_{\lambda\rightarrow 0}(\tilde{f})$ can be made by 
using the explicit form of $k(\tilde{f})=k_0e^{\tilde{f}}$. 
\begin{equation}
P_{\lambda\rightarrow\infty}(\tilde{f})=\frac{k_0\theta}{\tilde{\gamma}}\exp{\left[\tilde{f}-\frac{k_0\theta}{\tilde{\gamma}}(e^{\tilde{f}}-1)\right]}
\end{equation}
 and
 \begin{equation}
 P_{\lambda\rightarrow 0}(\tilde{f})=\frac{k_0\theta}{\tilde{\gamma}}\exp{(\tilde{f})}\left[1+2\frac{k_0\theta}{\tilde{\gamma}}(e^{\tilde{f}}-1)\right]^{-3/2}.
 \label{eqn:oneD1}
 \end{equation}
 
For $\tilde{f}\rightarrow\infty$,  
$P(\tilde{f})$ behaves as 
\begin{align}
\lim_{\tilde{f}\rightarrow \infty}\log{P_{\lambda\rightarrow\infty}(\tilde{f})}&\sim\tilde{f}-\frac{k_0\theta}{\tilde{\gamma}}\exp{(\tilde{f})}\nonumber\\
\lim_{\tilde{f}\rightarrow \infty}\log{P_{\lambda\rightarrow 0}(\tilde{f})}&\sim -\tilde{f}/2. 
\label{eqn:oneD2}
\end{align}
It is worth noting that depending on the $\lambda$ value ($\lambda\rightarrow \infty$ or 0) $P_{\lambda}(\tilde{f})$ differs in its asymptotic behavior with respect to $\tilde{f}$ (see Eqs.\ref{eqn:oneD1} and \ref{eqn:oneD2}). 

The asymptotic behavior of the so-called microscopic model \cite{Garg95PRB,Dudko06PRL}, whose force range is limited by the critical force ($f<f_c=\Delta G^{\ddagger}/\nu\Delta x^{\ddagger})$, is reduced to that of Gumbel distribution only if $f^*<f\ll f_c$. 
If $f^*<f\rightarrow f_c$ then the unbinding force distribution decays precipitously to zero as $\sim (1-f/f_c)^{1/\nu-1}$($\nu=2/3$: cubic potential) and linearly ($\nu=1/2$: harmonic cusp potential) ($\lambda\rightarrow\infty$ corresponds to the Bell model).
Note that the model in \cite{Garg95PRB,Dudko06PRL} corresponds to $\lambda\rightarrow\infty$.    

In contrast, for $\tilde{f}\rightarrow 0$, 
\begin{align}
\lim_{\tilde{f}\rightarrow 0}P_{\lambda\rightarrow\infty}(\tilde{f})&\sim \frac{k_0\theta}{\tilde{\gamma}}\times\exp{\left[\left(1-\frac{k_0\theta}{\tilde{\gamma}}\right)\tilde{f}\right]}\nonumber\\
\lim_{\tilde{f}\rightarrow 0}P_{\lambda\rightarrow 0}(\tilde{f})
&\sim \frac{k_0\theta}{\tilde{\gamma}}\times\left[1+\left(1-3\frac{k_0\theta}{\tilde{\gamma}}\right)\tilde{f}+\mathcal{O}(\tilde{f}^2)\right].
\end{align}
The initial slope of $P(\tilde{f})$ is determined by the value of $k_0\theta/\tilde{\gamma}$.

\end{document}